\documentclass[aps,prb,preprint,groupedaddress]{revtex4-1}  
\def\be{\begin{eqnarray}}     
\def\ee{\end{eqnarray}}  
\usepackage{graphicx}  
\def\Jcm{J/cm$^2$~}

\def\Wcm{W/cm$^2$~}
\usepackage{color}
  
\begin{document}  
  
\title{Time-dependent density functional theory of high-intensity,
short-pulse laser irradiation on insulators}
 
\author{S.A. Sato}
\affiliation{Graduate School of Pure and Applied Sciences, University of
Tsukuba, Tsukuba 305-8571, Japan}

\author{K. Yabana}
\affiliation{Center for Computational Sciences, University of Tsukuba,
Tsukuba 305-8577, Japan}
\affiliation{Graduate School of Pure and Applied Sciences, University of
Tsukuba, Tsukuba 305-8571, Japan}

\author{Y. Shinohara}
\affiliation{Max-Planck-Institut f\"ur Mikrostrukturphysik, Weinberg 2, D-06120 Halle, Germany}

\author{T. Otobe}
\affiliation{Advanced Photon Research Center, JAEA, Kizugawa, Kyoto
619-0615, Japan}

\author{K.-M.~Lee}
\affiliation{Center for Relativistic Laser Science, Institute for Basic Science, Gwangju 500-712, South Korea}

\author{G.F. Bertsch}
\affiliation{Department of Physics and Institute for Nuclear Theory,
University of Washington, Seattle 98195, U.S.A.}

%===========================================================  
\begin{abstract} 
We calculate the energy deposition by very short laser pulses in
SiO$_2$ ($\alpha$-quartz) with a view to establishing systematics 
for predicting damage and nanoparticle production. The theoretical
framework is time-dependent density functional theory, 
implemented by the real-time method in a multiscale representation.  
For the most realistic simulations we employ a meta-GGA Kohn-Sham 
potential similar to that of Becke and Johnson.  
We find that the deposited energy in the medium can be accurately 
modeled as a function of the local electromagnetic pulse fluence.
The energy-deposition function can in turn be quite well fitted to 
the strong-field Keldysh formula for a range of intensities from below
the melting threshold to well beyond the ablation threshold.
We find reasonable agreement between the damage threshold and the
energy required to melt the substrate.  
The ablation threshold estimated by the energy
to convert the substrate to an atomic fluid is higher than the measurement, 
indicating significance of nonthermal nature of the process. 
A fair agreement is found for the depth of the ablation.
\end{abstract}  
\maketitle  
%===========================================================  
%\section{introduction}  
%===========================================================  
 
{\it Introduction.} 
Ablation of surfaces by intense femtosecond laser pulses is potentially
a useful tool for material machining as well as preparation of nanosized
particles\cite{ci96,st96,li97,le98,ti99,le00,su02,am05,ga08,ch11,ut11,ba13}.  
It is important to have good models of the energy 
deposition and the ablation process to set up the pulse protocols for the
purpose in mind.  However, it is not easy to model the laser-target
interaction because one has to deal with the quantum physics at the atomic 
scale together with the pulse propagation at a mesoscopic scale.

There is a large literature on the modeling of the laser-target 
interaction dynamics \cite{st96,ka00,pe03,re041,re04,pe05,ch09,ch11,br14}, 
as reviewed in Ref. \cite{ga11} as well as Ref. \cite{ba13}.  
The fundamental physics is the excitation of 
particle-hole pairs, often parameterized by Keldysh's approximate 
formulas \cite{ke65,ti99}.
The particle and hole carriers affect the electromagnetic response of the 
insulator, screening the field when their density becomes large.  
Here the effects are
often parameterized by hybrid dielectric functions containing contributions
intrinsic to the insulator and plasma contributions from the excited
electrons.  Both aspects of this electron dynamics are included in the
time-dependent density functional theory (TDDFT), which is fully quantum
mechanical and doesn't require any specific assumptions about the dynamics.
It also gives a good compromise
between ab initio theory and computational feasibility.  In this
work we shall apply the dynamic equations of TDDFT to calculate
the propagation of short, intense electromagnetic pulses upon
insulators and compute the energy transfer to the medium. 

In recent work, TDDFT with strong fields has been applied to 
condensed media in several contexts:
coherent phonon generation \cite{sh10,sh12} and high-field interactions 
with diamond, silicon and quartz \cite{ot08,ot09,ya12,sa14,le14,wa14}.  
In this work, we treat the energy transfer to a crystalline SiO$_2$,
$\alpha$-quartz, a material of technological importance.  
The calculations require high-performance computers, so our goal
is to also analyze the results in terms of simpler models that
can be easily applied.  We only discuss the
deposited energy of the pulse, which is the most important
determinant of the subsequent atomic dynamics.
Since TDDFT employing practically usable functionals
does not have any relaxation processes, it is limited to short-time
dynamics.  The main limitations to consider are the electron-electron
kinetic relaxation time $\tau_{ee}$, the full electron thermalization time,
and the electron-phonon equilibration time.  The latter two processes 
have time scale much larger than 100 fs, which is well beyond the
time domain considered here.  Estimates of $\tau_{ee}$ range from 
1 fs to 100 fs \cite{ch11,ut11,br87,fa92}.  Fortunately for the
modeling by TDDFT, the response of insulators to high fields is
rather insensitive to the kinetic equilibration \cite{sa14b}.  
We note that on very long time scales another mechanism 
that is missing from TDDFT becomes important, namely 
an avalanche of electrons in the free carrier bands \cite{re04}.

{\it Computational method.} 
The theory and its implementation 
used in the present calculation
has been described elsewhere \cite{ya12}, 
so we will just summarize the major points.  
The dynamics are derived from a Lagrangian that
includes the full electromagnetic gauge field $\vec A(\vec{r},t)$ 
as well as the Kohn-Sham dynamics of the TDDFT.  
Thus the equations of motion can describe
the pulse propagation from the vacuum into the medium and 
its attenuation by exciting electrons from the valence 
to the conduction band.  The quantities of interest vary
on a scale of micrometers, while the electron dynamics requires
a sub-nanometer scale.  This conflict is overcome by 
separating contributions to the electric field into a contribution 
$\partial \vec A /c\partial t$ varying only on the larger scale and a 
contribution $-\vec \nabla \phi$ that has the full atomic-scale 
variation in the crystalline unit cell.
The coupling between the scales arises from the 
current $\vec J$ associated with the electrons in the unit cell.   

In this work we use a modified Becke-Johnson exchange-correlation 
potential (mBJ) \cite{be06b} as given by Ref. \cite[Eq. 2-4]{tr07}.  
The calculated indirect band gap with this potential is 1.3 eV lower 
than the experimental gap of $8.9$ eV, but more relevant
to electromagnetic interactions is the optical gap.
This is calculated to be $~9$ eV from the absorptive part
of the dielectric function for the mBJ potential, 
compared to the experimental value of $9-10$ eV.
We compute the energy transfer to the 
medium from the electromagnetic
side, since the calculation from the Kohn-Sham densities would require
an explicit energy density functional for the mBJ potential.
The energy transfer rate $W$ is given by  
\be
W = - \vec{\cal E} \cdot \vec J,
\ee
where $\vec {\cal E}$ is the electric field associated with the pulse.  
In comparing different pulse lengths and intensities, we find the most
convenient measure of the pulse strength is its fluence $F$
given by
$
F(x) = (c/4\pi) \int dt \hat{x} \cdot \vec{E} \times \vec{B},
$
where $x$ is the depth from the surface.
With these definitions, the deposited energy density
is given by 
$
E_x(x) = -\int dt W(x,t) = -{d \over d x} F(x).
$

Our multiscale calculation
uses a one-dimensional grid with spacing of $250$ au
for propagation of laser electromagnetic fields.
At each grid point, electron dynamics is calculated using atomic-scale 
rectangular unit cell containing 
$6$ Silicon atoms and $12$ Oxygen atoms which are discretized
into Cartesian grids of $20 \times 36 \times 50$.
The dynamics of the $96$ valence electrons is treated explicitly; 
the effects of the core electrons are taken into account by 
pseudopotentials  \cite{kl82}.  
Both electromagnetic fields and electrons
are evolved with a common time step of 0.02 au.

The laser pulse in the vacuum is described by a gauge field
$\vec A(x,t)$ of the form
\be
\vec A(x,t) = \hat z A_0\sin^2(\pi t_x/T_{p}) \cos(\omega t_x)
\ee
in the domain $0<t_x<T_p$ and zero outside.  
Here $\omega$ is the photon's angular frequency,
$t_x = t - x/c$ describes the space-time dependence of the field,
and $T_p$ controls the pulse length.  It is related to 
the usual measure $\tau_p$ (full width at half-maximum intensity)
by $\tau_p = 0.364 T_p$. 
Most of the results below were calculated for an average 
photon energy of $\hbar \omega = 1.55$ eV and the pulse
length of $\tau_p=7$ fs.
We follow the pulse from the time its front reaches
the surface ($\bar210$) until it has propagated several 
$\mu$m into the material.

\begin{figure}     
\includegraphics [width = 0.9\columnwidth]{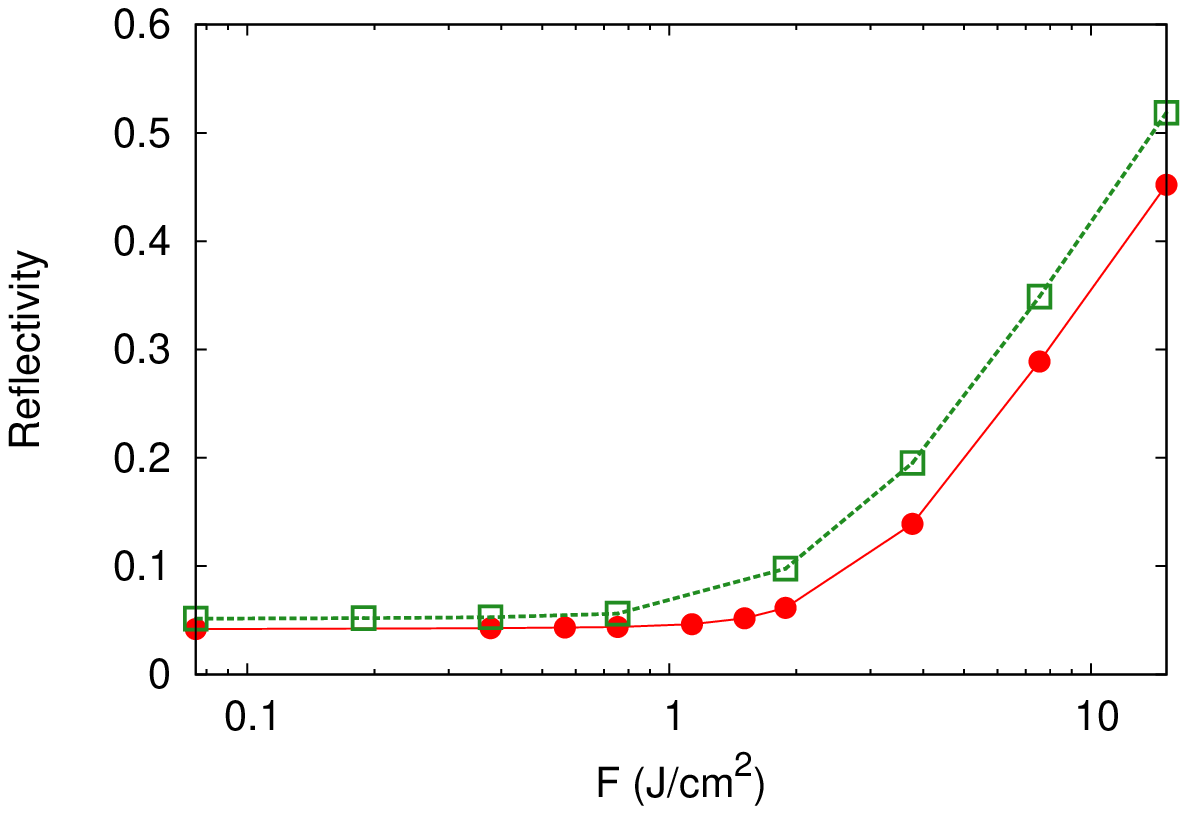}     
\caption{\label{R}   
The reflectivity at the surface of $\alpha$-quartz as a function 
of fluence. Red circles show results using mBJ potential, while
green squares show results using the LDA functional 
taken from Ref. \cite{le14}.}
\end{figure}

{\it Results.}
We first note that our calculated reflectivity of $\alpha$-quartz, 
shown in Fig. \ref{R},  is close to $4\%$ for laser
fluence below 1.5 J/cm$^2$ increasing to 
about $35\%$ at fluence 10 J/cm$^2$.
The rise at high field intensity is due to the screening effect 
of an electron-hole plasma that is formed.  
The Figure shows the reflectivity for both the
mBJ potential and the LDA energy functional from Ref. \cite{le14}.    
One may see that the screening has a lower threshold for LDA functional.
This can be attributed to the lower band gap in LDA, permitting
a more dense plasma to form.

The main quantity we can calculate is the deposited energy $E_x$ as a
function of penetration depth $x$.   
The range of interest extends roughly from the energy required to melt the
solid to the energy required to vaporize it.  The first transition
requires about 0.5 eV/atom in quartz, starting from room temperature.
The second transition is not as well defined; we can estimate
it as 6 eV/atom either as the heat of formation or from 
the ``atomic-liquid" transition reported in Ref. \cite{hi06}.

\begin{figure}     
\includegraphics [width = 0.9\columnwidth]{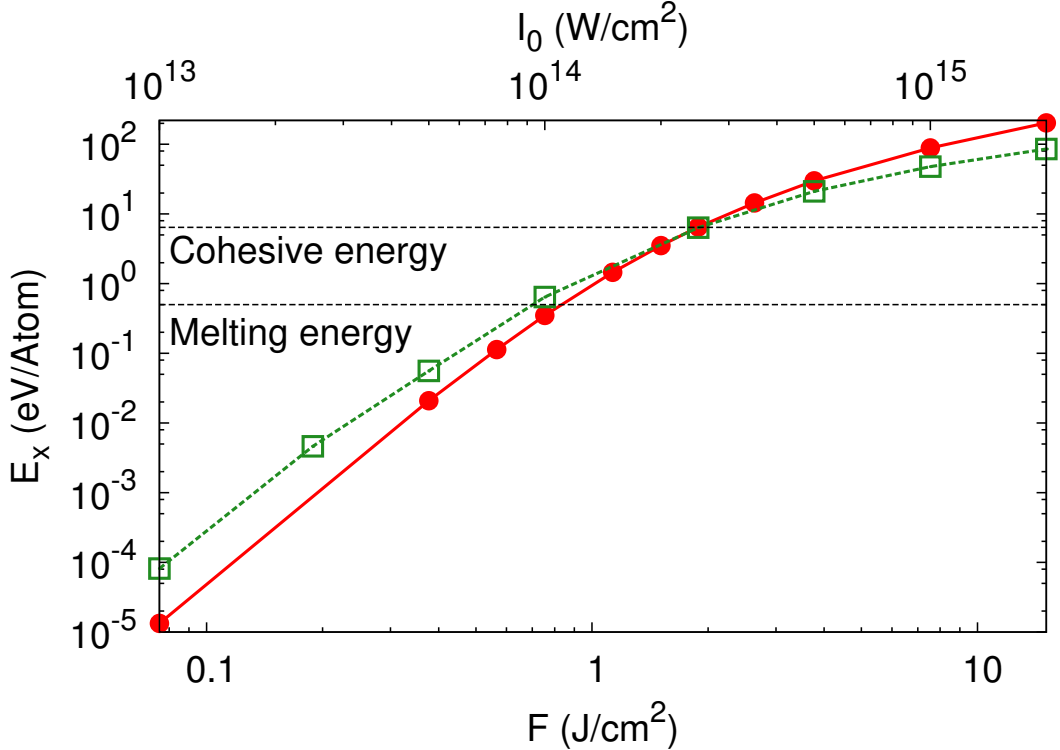}     
\caption{\label{surface}   
Deposited energy $E_x$ at the surface of $\alpha$-quartz is shown, 
as a function of the fluence and the peak intensity of the pulse. 
Red circles show results using mBJ potential, 
while green squares show results using LDA functional \cite{le14}. }  
\end{figure}

A sampling of the results for the multiscale calculation is shown in
Figs. 2-7.  In Fig. \ref{surface}, we show the deposited energy 
density at the surface for a range of laser fluences.  
The threshold for achieving melting is $\sim 1.5$ \Jcm, while the 
vaporization threshold as we have defined it is at $\sim 2.5$ \Jcm.
Thus, for applications that physically transform the material,
we need to only consider fluences of the order $1.5$ \Jcm and 
and higher.  The Figure shows the deposited energies of lower
strength pulses as well, which may be of interest to analyze the 
parametric dependence of the energy deposition on the 
characteristics of the pulse. 
The Figure also shows the results for the LDA functional.  
One sees that lower-fluence pulses are more highly absorbed 
in LDA .  This is to be expected due to the small band gap.  
The difference becomes small and even changes sign at high fluence.  
A possible explanation might be an increased screening for the LDA, 
as was discussed for the reflectivity. 

\begin{figure}     
\includegraphics [width = 0.9\columnwidth]{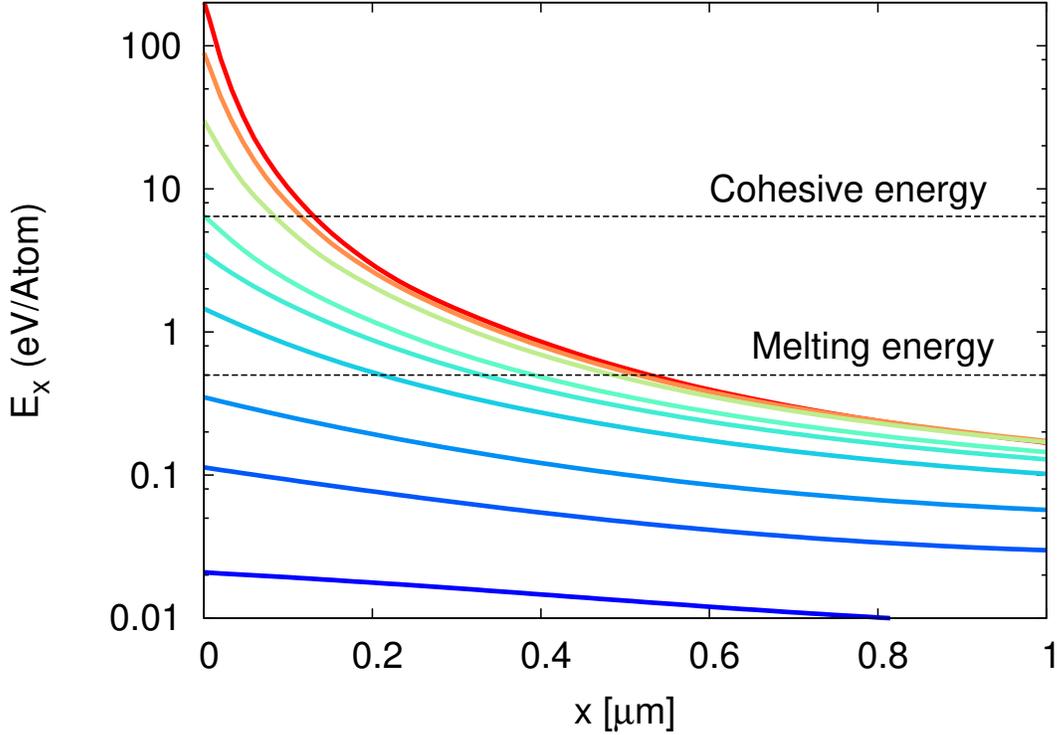}     
\caption{\label{EvsI}   
Deposited energy $E_x$ in $\alpha$-quartz as a function of 
distance below the surface.  
The curves show the results for intensities
$I_0 = 0.5,0.75,1.0,1.5,2.0,2.5,5.0,10.0,20.0 \times 10^{14}$ \Wcm
from lowest to highest graph.  The pulse length is 
$\tau_p = 7$ fs for all intensities.
}  
\end{figure}

The dependence of $E_x$ on the depth in the medium 
is shown in Fig. \ref{EvsI} for a range of pulse intensities, 
all with pulse length of $\tau_p  = 7$ fs.
One sees that the melted region extends to a depth of 0.5-0.6
$\mu$m for the stronger pulses.  The depth permitting ablation only
extends to 0.13 $\mu$m for a pulse of 10 times the threshold 
intensity.

\begin{figure}     
\includegraphics [width = 0.9\columnwidth]{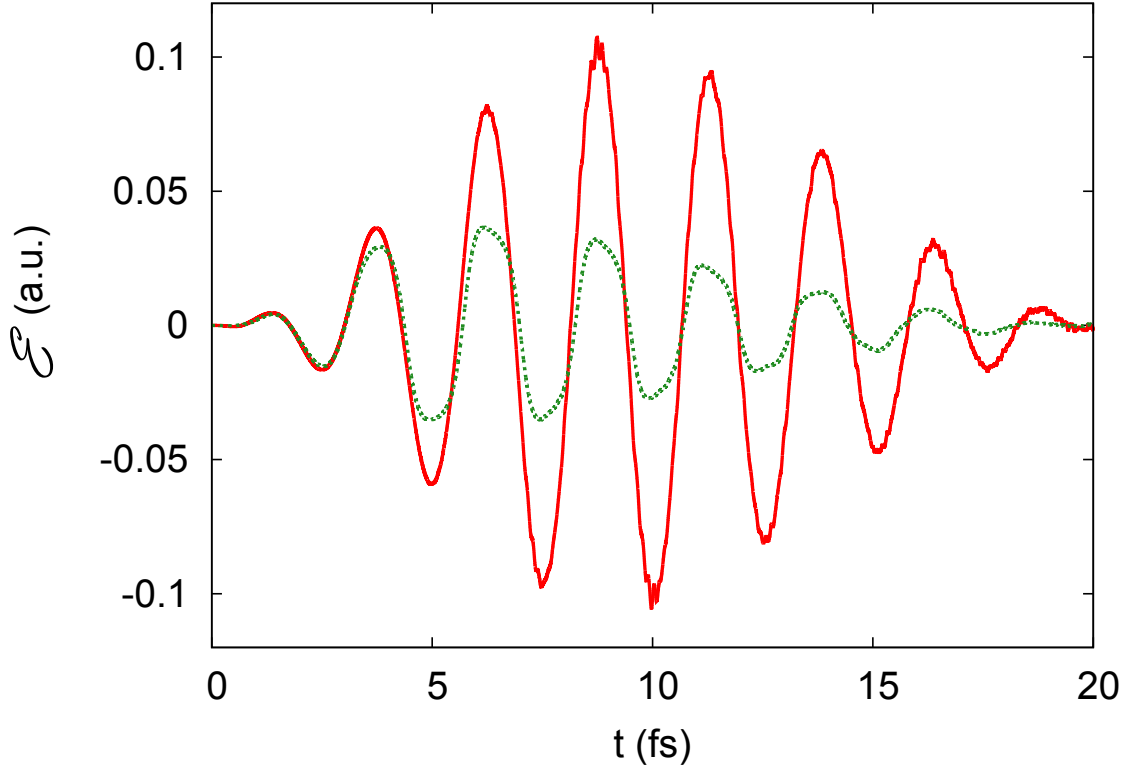}     
\caption{\label{pulse-shape}   
Pulse shapes at the first cell and at the 80th cell located
at a depth of ~1 $\mu$m are shown, for an incident pulse 
having intensity $I_0 = 10^{15}$ \Wcm and length $\tau_p = 7$ fs.}  
\end{figure}

Fig. \ref{pulse-shape} compares the pulse profile for the $I_0 =  10^{15}$ \Wcm pulse
at two points, in the first cell and in the 80th cell at a depth
of 1.06 $\mu$m.  At this depth, the
intensity has decreased below the melting threshold. One
sees that the shapes are quite similar.  The main difference is
that the attenuation is stronger in the later arriving cycles
of the pulse train.  This is to be expected; the early part of 
the pulse creates particle-hole pairs which can then modify
the propagation of the rest of the pulse.

\begin{figure}     
\includegraphics [width = 0.9\columnwidth]{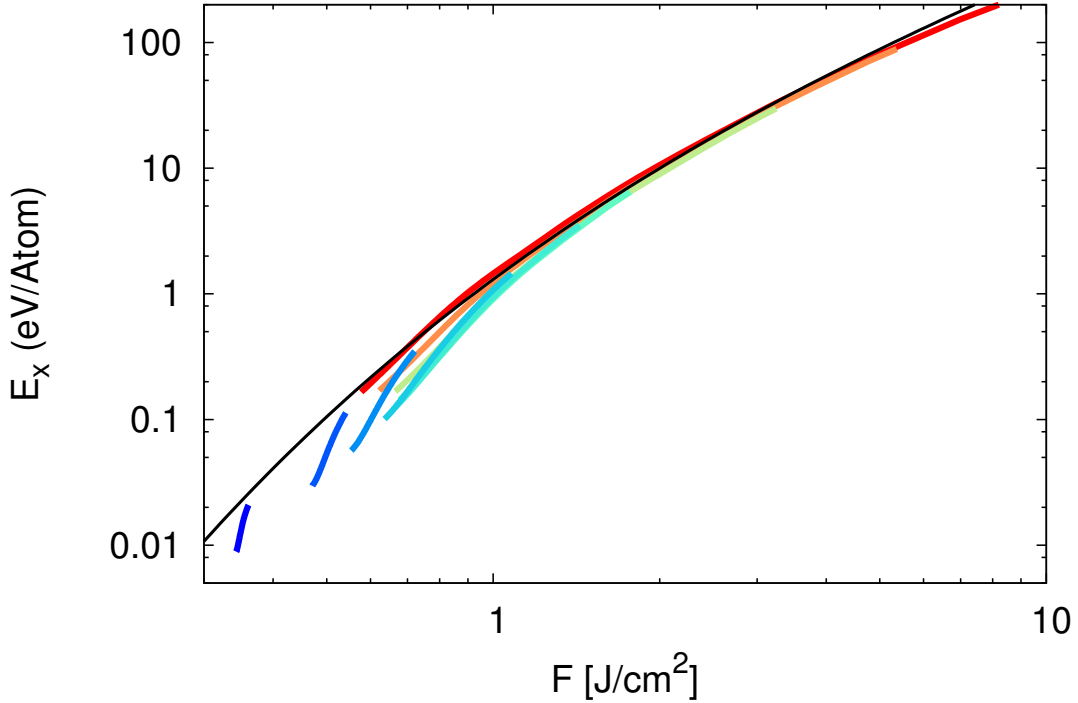}     
\caption{\label{Ea-vs-F}   
Absorbed energy is plotted as a function of local fluence of pulses.
Black solid curve shows a fit by Keldysh's formula.
}  
\end{figure}

{\it Approximate description.}  
The mild changes in pulse shape suggest that the energy 
deposition might be modeled simply as a function of the strength of the 
pulse as it is attenuated in the medium.  To see how well this
works, we extract the local fluence of the pulses at the 
different cells in our simulation.  The absorbed energy versus
the local fluence for the range of pulses shown in Fig. \ref{EvsI} is
plotted in Fig. \ref{Ea-vs-F}.  We see that they fall on a common line,
extending down to a fluence of ~1 \Jcm, well below the 
damage threshold.  
Thus, the change in pulse shapes can indeed be 
neglected, at least for $t_p = 7$ fs pulses.
Remarkably, the dependence can be very well fitted following 
Keldysh's strong-field ionization rate equation\cite[Eq. (9)]{ti99}.  
An approximate formula expressing the energy deposition 
in terms of the fluence is
\be
E_x = A F^{5/4} \exp( - B/F^{1/2}).
\ee
The two-parameter fit ($A = 70$  eV-(\Jcm)$^{-5/4}, $B= 4.0$
$(\Jcm)$^{1/2}$)
is shown  as the solid black line in the Figure.
One sees that the fit is valid from fluences 
from well below the thresholds to the highest calculated.  
The fit value $B=4.0$ can be compared to the value obtained from a
reduction of Keldysh's exponential factor,
\be 
\label{exponential}
B = \pi \tau_p^{1/2}m^{1/2} \varepsilon^{1/4}\Delta^{3/2} /2,
\ee
in atomic units.
Taking the reduced mass $m=1/2$, the direct band gap $\Delta = 9$ eV, 
and the dielectric constant $\varepsilon = 2.3$, Eq. (\ref{exponential})
gives  $B=4.2$, a difference of only 5\% from the fit value.
The multiphoton process is only relevant
at much lower fluences that are needed for structural changes.

\begin{figure}     
\includegraphics [width = 0.9 \columnwidth] {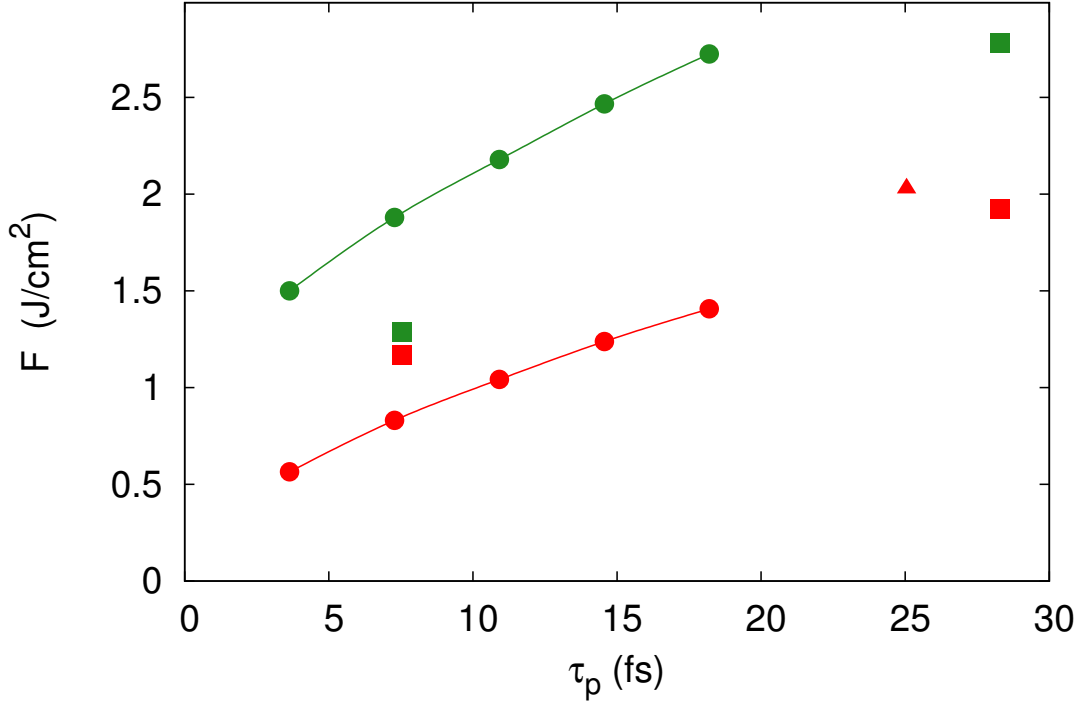}
\caption{\label{data}  
Threshold damage and threshold ablation fluences in SiO$_2$ are
shown as a function of pulse width $\tau_p$.  
Red and green circles are the calculated fluences 
to deposit an energy $E_x = 0.5$ eV/atom  corresponding
to a damage threshold and an energy $E_x = 6$ eV/atom 
corresponding to an ablation threshold in $\alpha$-quartz,
respectively.
Experimental data from Ref. \cite{ch11} for fused silica 
is shown with green and red squares; data from Ref. \cite{ti99} 
is shown with red triangle.
}
\end{figure}

{\it Experimental.}  While it is not entirely clear how energy deposition
profiles link to structural changes in the surface region, we can still
compare the theory and experiment assuming that the melting and 
vaporization transitions control the surface damage and ablation.
There are many measurements of thresholds for these quantities, 
but only a few for pulse as short as $\tau_p \sim 20$ fs or below \cite{le98,ti99,le00,ch11}.  
Fig. \ref{data} shows a comparison with the data of 
Refs. \cite{ti99,ch11} for fused silica.
For the damage threshold shown by red symbols, we compare 
with calculated fluences to achieve $E_x = 0.5 $ eV/atom at 
the surface. We see that the experimental threshold is in 
qualitative correspondence with that value.
The ablation threshold, shown by green symbols, is much lower
than predicted by the $E_x = 6$ eV/atom criterion based on the
formation of an atomic liquid \cite{hi06}. 
This may indicate an importance of nonthermal effects
in the ablation process by femtosecond laser pulses
\cite{lo03,ku14}.
In fact, the reported thresholds for damage and ablation at 
$\tau_p = 7 $ fs in Ref. \cite{ch11} are nearly identical.  
This may be seen by the sharp edges of the ablation
craters formed at the shortest pulse lengths.
Perhaps one needs to consider a different mechanism for ablation at
threshold, which may involve the electric fields produced when
excited electrons are ejected from the surface.
It should also be kept in mind that damage threshold depends on
structures of the material; a lower threshold is reported for
fused silica than crystalline SiO$_2$ using much longer pulses\cite{xu07}.

\begin{figure}     
\includegraphics [width = 0.9 \columnwidth]
{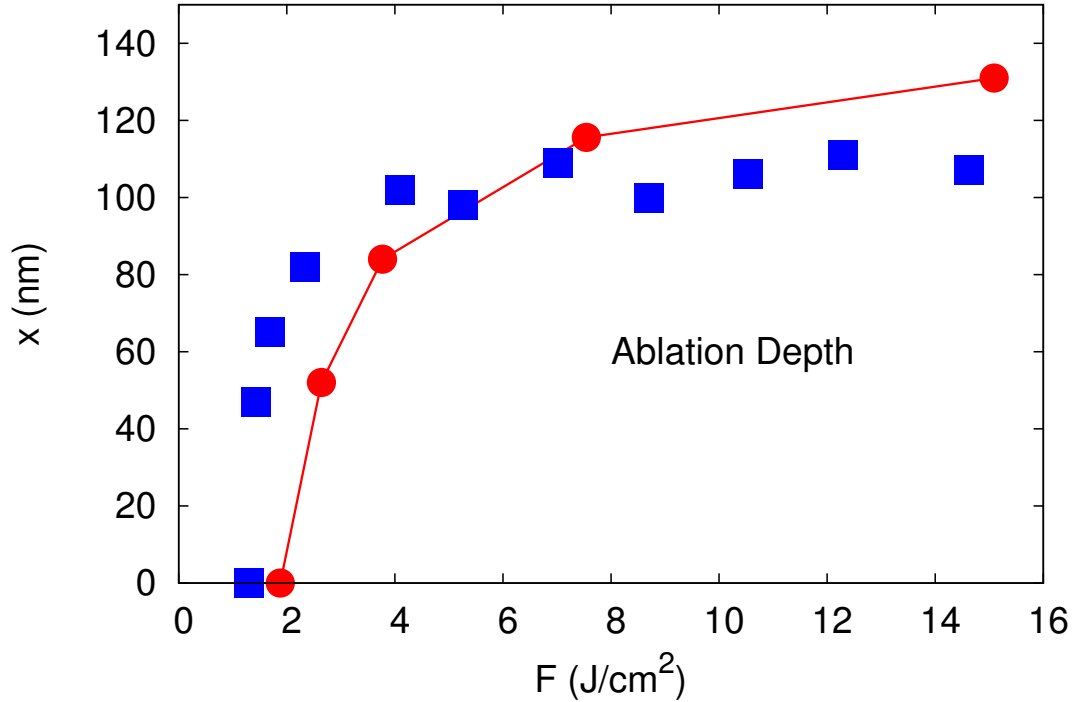}
\caption{\label{depth}   
Depth of crater formed in SiO$_2$ by single laser pulses of length
$\tau_p = 7 $, as a function of the fluence of the original pulse.   
Experiment is shown in blue squares, from Ref. \cite{ut11}.  The red circles 
show the depth at which the calculated
energy deposition falls below $E_x = 6$ eV/atom.  
}  
\end{figure}

As a final theory-experiment comparison,  we examine the depth of the 
ablated craters as a function of the fluence of the pulses.  
The available experimental data is shown
in Fig. \ref{depth}.  For the theory, we report the depth at 
which the deposited energy falls to $E_x = 6$ eV/atom, as in
Fig. \ref{data}.  The agreement between
theory and experiment is quite satisfactory.  The theory reproduces the
very sharp rise above threshold as well as the saturation at high
fluences.  

{\it Conclusion.} We have shown that it is feasible to calculate the
interaction of intense femtosecond laser pulses with insulating media
by the TDDFT, avoiding the detailed modeling of plasma formation and 
dynamics required in earlier theoretical treatments.  The threshold
for damage is accounted for by the calculated energy deposition needed
to melt the quartz substrate.  On the other hand, the calculated threshold
for ablation via a transformation to an atomic liquid is 50\% higher than 
the low values observed in two experiments.  
This may suggest the significance of nonthermal mechanism 
in the ablation process. It may also indicate that another mechanism is 
responsible for the threshold behavior, such as an electric-field assisted ionic dissociation.  However, we find good agreement with experiment on the
depth of the ablation taking the energy deposition criterion to estimate
the depth.  It may well be that modeling energy deposition in wide-gap 
insulators could be greatly simplified once one has a suitable set of 
benchmark multiscale calculations based on the TDDFT.

%===========================================================  
\section*{Acknowledgments}

This work was supported by JSPS KAKENHI Grant Numbers
23340113, 25104702,  26-1511.
This work used computational resources of the K computer provided by the
RIKEN Advanced Institute for Computational Science through the HPCI
System Reserch project (Project ID: hp140103).
G.F.B. acknowledges support 
by the US Department of Energy under Grant No. DE-FG02-00ER41132.


\begin{thebibliography}{99}  
\bibitem{ci96} B.N. Chichkov, C. Momma, S. Nolte, V. von Alvensleben, 
A. Tunnermann, Appl. Phys.  A{\bf 63}, 109 (1996).
\bibitem{st96} B.C. Stuart, M.D. Feit, S. Herman, A.M. Rubenchik, 
B.W. Shore, M.D. Perry, Phys. Rev. B{\bf 53}, 1749 (1996).
\bibitem{li97} X. Liu, D. Du, G. Mourou, IEEE J. Quantum Electronics {\bf 33}, 
1706 (1997).
\bibitem{le98} M. Lenzner, J. Krugeer, S. Sartania, Z. Cheng, C. Spielmann, 
G. Mourou, W. Kautek, F. Krausz, Phys. Rev. Lett. {\bf 80} 4076 (1998).
\bibitem{le00} M.~Lenzner, F.~Krausz, J.~Kr\"uger, and W.~Kautek,
Applied Surface Science {\bf 154} 11 (2000).
\bibitem{su02} L. Sudrie, A. Couairon, M. Franco, B. Lamouroux, B. Prade, 
S. Tzortzakis, A. Mysyrowicz, Phys. Rev. Lett. {\bf 89}, 186601 (2002).
\bibitem{am05} S. Amoruso, G. Ausanio, R. Bruzzese, M. Vitiello, X. Wang,
Phys. Rev. B{\bf 71}, 033406 (2005).
\bibitem{ga08} R.R. Gattass, E. Mazur, Nature Photonics {\bf 2}, 219 (2008).
\bibitem{ch11}B. Chimier, et al., Phys. Rev. B {\bf 84} 094104 (2011).
\bibitem{ut11} O.~Ut\'eza, et al., Appl. Phys. A {\bf 105} 131 (2011).
\bibitem{ba13} P. Balling and J. Schou, Rep. Prog. Phys. {\bf 76}
036502 (2013).
\bibitem{ka00} B. Kaiser, B. Rethfeld, M. Vicanek, G. Simon,
Phys. Rev. B{\bf 61}, 11437 (2000).
\bibitem{pe03} D. Perez, L.J. Lewis, Phys. Rev. B{\bf 67}, 184102 (2003).
\bibitem{re041} C. Rethfeld, K. Sokolowski-Tinten, D. von der Linde, S.I. Anisimov,
Appl. Phys. A{\bf 79}, 767 (2004).
\bibitem{re04} B.~Rethfeld, Phys. Rev. Lett. {\bf 92} 187401 (2004).
\bibitem{pe05} J.R. Penano,  P. Sprangle, B. Hafizi, W. Manheimer, A. Zigler,
Phys. Rev. E{\bf 72}, 036412 (2005).
\bibitem{ch09} B.H. Christensen, P. Balling, Phys. Rev. B{\bf 79}, 155424 (2009).
\bibitem{br14} N. Brouwer, B. Rethfeld, J. Opt. Soc. Am. B{\bf 31}, C28 (2014).\bibitem{ga11} E.G. Gamaly, Phys. Rep. {\bf 508}, 91 (2011). 
\bibitem{ke65} L.V.~Keldysh, Sov. Phys. JETP {\bf 20} 1307 (1965).
\bibitem{ti99} A.-C.~Tien, S. Backus, H. Kapteyn, M. Murname, G. Mourou, 
Phys. Rev. Lett. {\bf 82} 3883 (1999).
\bibitem{sh10} Y.~Shinohara, K.~Yabana, Y.~Kawashita, J.-I.~Iwata,
T.~Otobe,  G.F.~Bertsch, Phys. Rev. {\bf B82} 155110 (2010).    
\bibitem{sh12} Y. Shinohara, S.A. Sato, K. Yabana, T. Otobe, J.-I. Iwata, G.F. Bertsch, 
J. Chem. Phys. {\bf 67}, 22A527 (2012).
\bibitem{ot08} T. Otobe, M. Yamagiwa, J.-I. Iwata, K. Yabana, T. Nakatsukasa, G.F. Bertsch,
Phys. Rev. B{\bf 77}, 165104 (2008).
\bibitem{ot09} T.~Otobe, K.~Yabana, and J.-I.~Iwata, J. Phys. Cond. Matt.
{\bf 21} 064224 (2009).
\bibitem{ya12} K. Yabana, T. Sugiyama, Y. Shinohara, T. Otobe, G.F. Bertsch,
Phys. Rev. B {\bf 85} 045134 (2012). 
\bibitem{sa14} S.A.~Sato, K. Yabana, Y. Shinohara, T. Otobe, G.F. Bertsch, 
Phys. Rev. B {\bf 89} 064304 (2014).
\bibitem{le14} K-M Lee, C.M. Kim, S.A. Sato, T. Otobe, Y. Shinohara, 
K. Yabana, and T.M. Jeong, J. App. Phys. {\bf 115} 053519 (2014).  
\bibitem{wa14} G.~Wachter, C. Lemell, J. Burgd\"orfer, S.A. Sato, X.-M. Tong, K. Yabana, 
Phys. Rev. Lett. {\bf 113} 087401 (2014).
\bibitem{br87} S.D.~Brorson, J.G.~Fujimoto, and E.P.~Ippen, Phys. Rev.
Lett. {\bf 59} 1962 (1987).
\bibitem{fa92} W.S.~Fann, R. Storz, H.W.K. Tom, J. Bokor, 
Phys. Rev. B {\bf 46} 13592 (1992).
\bibitem{sa14b} S.A.~Sato, Y.~Shinohara, T.~Otobe and K.~Yabana, 
Phys. Rev. B {\bf 90}, 174303 (2014).
\bibitem{be06b} A.D.~Becke and E.R.~Johnson, J. Chem Phys. {\bf 124}
221101 (2006).
\bibitem{tr07} F. Tran and P. Blaha, Phys. Rev. Lett. {\bf 102}, 226401 (2009). 
\bibitem{kl82} L. Kleinman and D. M. Bylander,  
Phys. Rev. Lett. {\bf 48}, 1425 (1982). 
\bibitem{hi06} D.G.~Hicks, et al., Phys. Rev. Lett. {\bf 97} 025502 (2006).
\bibitem{lo03} P. Lorazo, L.J. Lewis, M. Meunier, 
Phys. Rev. Lett. {\bf 91}, 225502 (2003).
\bibitem{ku14} T. Kumada, H. Akagi, R. Itakura, T. Otobe, A. Yokoyama,
J. Appl. Phys. {\bf 115}, 103504 (2014).
\bibitem{xu07} S. Xu, J. Qiu, T. Jia, C. Li, H. Sun, Z. Xu, Opt. Comm. {\bf 274}, 163 (2007).

\end{thebibliography}
\end{document}